\documentclass[pra,aps,amsmath,twocolumn,showpacs,floatfix]{revtex4-1}
\usepackage{graphicx}
\usepackage{bm}
\usepackage{pdfpages}
\usepackage{color}
\usepackage[normalem]{ulem}
\input{epsf}
\begin{document}
\epsfverbosetrue
\def\la{{\langle}}
\def\ra{{\rangle}}
\def\vep{{\varepsilon}}
\newcommand{\beq}{\begin{equation}}
\newcommand{\eeq}{\end{equation}}
\newcommand{\beqa}{\begin{eqnarray}}
\newcommand{\eeqa}{\end{eqnarray}}
\newcommand{\q}{\quad}
\newcommand{\A}{\alpha}
\newcommand{\h}{\hat{H}}
\newcommand{\ha}{\hat{h}}
\newcommand{\p}{\partial}
\newcommand{\dg}{+}
\newcommand{\dga}{\dagger}
\newcommand{\AC}{{\it AC}}
\newcommand{\n}{\\ \nonumber}
\newcommand{\om}{\omega}
\newcommand{\Om}{\Omega}
\newcommand{\os}[1]{#1_{\hbox{\scriptsize {osc}}}}
\newcommand{\cn}[1]{#1_{\hbox{\scriptsize{con}}}}
\newcommand{\sy}[1]{#1_{\hbox{\scriptsize{sys}}}}

\newcommand{\tcr}[1]{\textcolor{red}{#1}}
\newcommand{\tcb}[1]{\textcolor{blue}{#1}}
%


\title{Symmetry-assisted resonance transmission of non-interacting identical particles}
\author {D. Sokolovski$^{a,b}$}
\author {J. Siewert$^{a,b}$}
\author {L.M. Baskin$^c$}
\affiliation{$^a$ Departmento de Qu\'imica-F\'isica, Universidad del Pa\' is Vasco, UPV/EHU, Leioa, Spain}
\affiliation{$^b$ IKERBASQUE, Basque Foundation for Science, E-48011 Bilbao, Spain}
\affiliation{$^c$The Bonch-Bruevich State University of Telecommunications,
 193232, Pr. Bolshevikov 22-1, Saint-Petersburg, Russia}
\date{\today}

\begin{abstract}
We show that a ``pile up" effect occurring for a train of non-interacting identical particles 
incident on the same side of a resonance scatterer leads to significant
interference effects,
different from those observed in Hong-Ou-Mandel experiments.
These include characteristic 
changes in the overall transmission rate and full counting
statistics, as well as "bunching" and "anti-bunching" behavior in the all-particles transmission 
channel. With several resonances involved, pseudo-resonant
driving of the two-level system in the barrier, may also result in sharp 
enhancement of scattering probabilities for certain values of temporal delay between the particles.

\end{abstract}
\maketitle
\vskip0.5cm
\maketitle

%
%
%
%
%
\section {Introduction}
Quantum statistical effects accompanying scattering of non-interacting identical particles, are among some of the most intriguing predictions of quantum mechanics. 
Their studies, both theoretical and experimental, now constitute an extensive research field \cite{Aaronson2013}-\cite{Hassler2008}.
 If two such  particles, prepared in wave packet states meet head on in free space, they would eventually ``pass through each other", just like their distinguishable counterparts. The situation is different if such particles coincide inside a scatterer, with a possibility of two (or more) distinct scattering outcomes for each particle.
 In the celebrated Hong-Ou-Mandel (HOM) experiment~\cite{HOM1987}, 
the particles entering a scatterer from opposite sides, are seen to leave the barrier predominantly from the same side (bosons), or from the opposite sides (fermions). 
 The HOM effect
 has found important practical applications in quality testing of single-photon
sources~\cite{Sun2009}, entanglement detection~\cite{Shih1988}, 
entanglement swapping~\cite{Gisin2007}, 
and quantum metrology~\cite{Zeilinger2004}. Its generalizations, to name  a few, include observation of multiple photon bunching effects \cite{mfb1,mfb2,mfb3} scattering of photons by  multiport beam splitters and using them as interferometers  for identical particle in spatially separated modes \cite{TichyPRL2010,Mayer2011,SpagnoloPRL2013}, 
Efforts to extend the HOM  interference experiments to bosononic or fermionized cold atoms~\cite{TG} are currently underway~\cite{AspectPRA2013,Aspect2015}. 
 
Perhaps surprisingly, little studied to date remains the case complementary to that of  Hong, Ou and Mandel, in which the particles enter the barrier from the {\it same} side and ``meet" there owing to a ``pile up effect", if the barrier is capable of detaining the first particle long enough for the following particle(s) to catch up with it. The statistical effects are, in this case, quite different from those predicted for the HOM interference, and are most pronounced in resonance tunneling, where a particle spends in the barrier roughly the lifetime of the corresponding metastable state, which can be large for sharp resonances. 

In this paper, we give the general theory of the effect and demonstrate how the said ``piling up" alters the transmission rate for initially correlated many-particle states. There are complex interference effects in the scattering statistics of the incident particles, whose wave packet modes do not overlap prior to their arrival at the scatterer. A preliminary analysis of the case of two fermionized atoms can be found in \cite{Baskin2014}.
In \cite{CAT} it was demonstrated that  interference effects of the similar kind would arise whenever a particle simultaneously populates several wave packet modes.
For brevity we will refer as particles (fermions or bosons) to both 
cold atoms and photons, equally amenable to our analysis.
\newline
The rest of the paper is organised as follows: in Section II we discuss the correlation between initial particles. In Section III we analyse the correlations acquired in scattering. In Section IV we introduce a generatng function for the scattering statistics.  In Section V we show how quantum statistical effects vanish for particles well separated initially. Sections VI and VII discuss the two- and $N$-particle cases, respectively. In Section VIII we aplly our general approach to resonance transmission across a scatterer supporting one or more metastable states. Section IX contains our conclusions.
%
\begin{figure}
	\centering
		\includegraphics[width=8cm,height=3cm]{{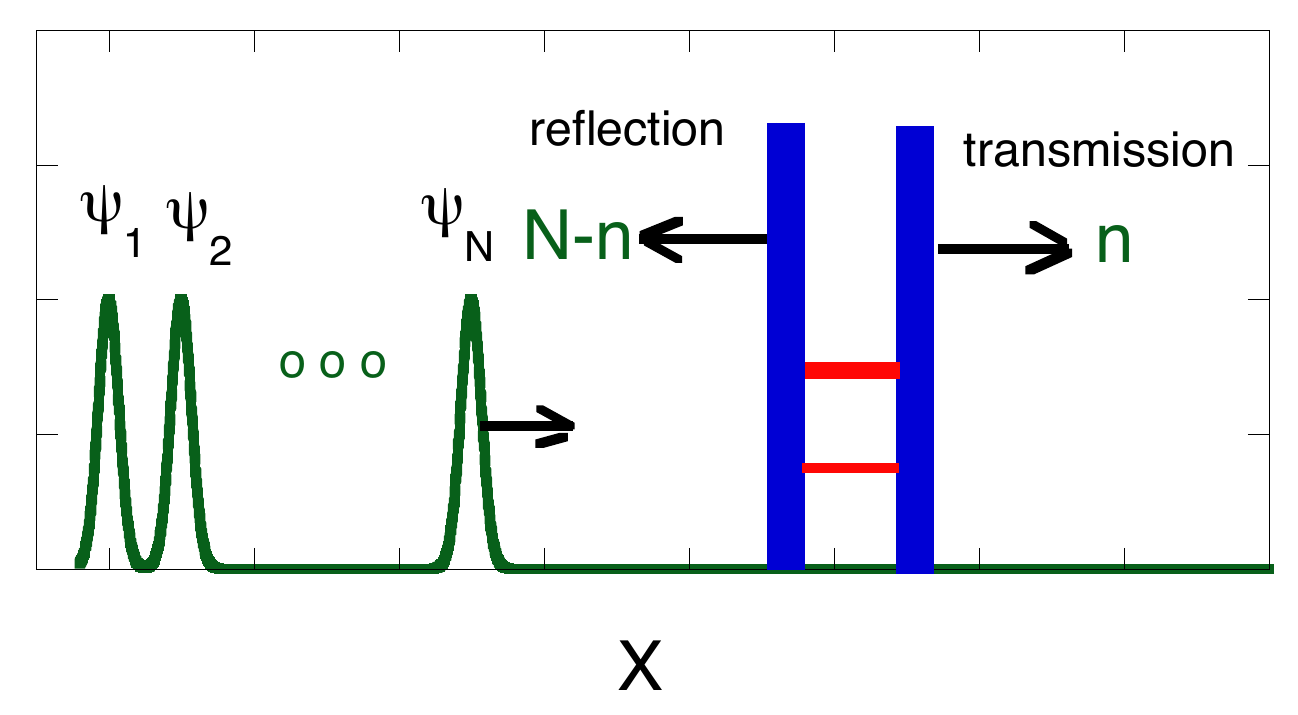}}
\caption{(Color online) Schematics diagram showing a "train" of $N$ wave packets incident on a double barrier which
supports two resonance levels.}
\label{fig:2}
\end{figure}
\section {Initial correlations between the particles}

Consider, in one dimension, a source sending $N$ identical noninteracting particles
of mass $\mu$~\cite{FOOTm}
 in wave packet states,
$\psi_n(x_n)$, as is illustrated in Fig.1,
\begin{eqnarray}\label{a1}
\nonumber
\psi_n(x_n,t)=(2\pi)^{-1/2}\int A_n(p) \exp[ipx_n-iE(p)(t+t_n)]dp
\ , \\ E(p)=p^2/2\mu\q\q
\end{eqnarray}
 towards a finite-width potential barrier at times $0=t_1<t_2 < ...< t_N$. 
\newline 
The operators $a_n^+$ and $a_n$, creating and annihilating an incident particle particle in the state $\psi_n$, are given by
\begin{eqnarray}\label{a2}
a_n^{\dg}= (2\pi)^{-1/2}\int A(p)\exp[-iE(t+t_n)] a^\dg(p) dp,\n
a_n=(a^\dg_n)^\dga,\q\q\q 
\end{eqnarray}
where the dagger denotes Hermitian conjugation,
 and the plane wave creation and annihilation operators $a^\dg(p)$ and $a(p)$ obey the usual commutation relations,  
$[a(p),a^\dg(p')]_{\mp}\equiv a(p)a^\dg(p')\mp a^\dg(p')a(p)=2\pi \delta(p-p'),$
with the upper and lower signs corresponding to bosons and fermions, respectively (cf.\ also Ref.~\cite{Loudon1998}). 
Their  (anti) commutators coincide with the overlaps between the 
wave packets (\ref{a1}),
\begin{align}\label{a4}
[a_m,a^\dg_n]_{\mp} \ =\ &  \int A^*_m(p)A_n(p)\exp[iE(p)(t_m-t_n)] dp
\nonumber\\
                 \equiv\ & I_{mn}\q\q
\end{align}
where  $I_{nn}=1$ and $I_{mn}=I^*_{nm}$
The symmetry of the incident state has no effect on the initial probability density provided  $I_{nm}=\delta_{nm}$,
e.g., for the delays between emissions, $|t_m-t_n|$, large enough for the
rapid oscillations of the exponential in (\ref{a4}) to destroy the integral for all $m\ne n$.
We will refer to such particles as {\it initially uncorrelated}. 
Alternatively, the particles may be prepared in a {\it correlated} initial state, and below we will consider both these cases.
It is readily seen that spreading of freely moving wave packets doesn't alter the commutation relations (\ref{a4}).
Thus, the symmetrized or anti-symmetrized wave function describing $N$ incident particles is given by
\begin{eqnarray}\label{a71}
|\Psi_{in}(t)\ra =K^{-1/2} \prod_{n=1}^Na^\dg_n(t)|0\ra,
\end{eqnarray}
where $|0\ra$ is the vacuum state, and $K$ is the normalization constant. 
By Wick's theorem,  
we have (the upper and lower signs are for bosons and fermions, respectively)
\begin{eqnarray}\label{a8}
K=\sum_{\sigma(N)}(\pm1)^{p(\sigma(N))}\prod_{i=1}^{N}I_{i\sigma_i}\equiv S^{\pm}[I_{mn}]
\end{eqnarray}
where $\sigma(N)$ is a permutation of the indices $(0,1,..,N)$ and $p(\sigma)$ is its parity \cite{FOOT1}.
\section {Correlations between the scattered particles}
At large times, after all particles have left the barrier area, each wave packet ends up split into the transmitted ($t$) and reflected ($r$) parts.
Thus, as $t\to \infty$,  the wave function has the form 
\begin{eqnarray}\label{a7}
|\Psi_{\text{out}}(t)\ra \ =\ 
      K^{-1/2} \prod_{n=1}^N[t_n^\dg(t)+r_n^\dg(t)]|0\ra
\ \ .
\end{eqnarray}
 where the  corresponding creation and annihilation operators are given by
%
\begin{eqnarray}\label{a5}
\nonumber
t^\dg_n= \int T(p)A_n(p)\exp[-iE(t+t_n)] a^\dg(p) dp \q ,\q\q\q \n
r^\dg_n= \int R(p)A_n(p)\exp[-iE(t+t_n)] a^\dg(-p) dp\q ,\q\q\\
t_n=(t^\dg_n)^\dg, \q r_n=(r^\dg_n)^\dga,\q\q\q
\end{eqnarray}
and  $T(p)$ and $R(p)$ are the barrier transmission and reflection amplitudes  for a particle with a momentum $p$.
Since $|T(p)|^2+|R(p)|^2=1$, as $t\to \infty$ we also have 
\begin{align}\label{a6}
\nonumber
T_{mn}\ \equiv\ & [t_m,t^\dg_n]_{\mp} 
\\     
            = \ &  
              \int |T(p)|^2A^*_m(p)A_n(p)\exp[iE(p)(t_m-t_n)] dp\ \ ,
\nonumber\\
 R_{mn} \equiv\ & [r_m,r^\dg_n]_{\mp} \ =\ I_{mn}-T_{mn},\q\q\q\q\q
\end{align}
while all remaining commutators vanish. 
In Eqs. (\ref{a6})  $T_{mn}=T^*_{nm}$ is a Hermitian matrix of the overlaps between the transmitted parts of the wave packets, 
and its diagonal elements $T_{nn}$ coincide with the probabilities
$w_n$ for the $n$-th particle to be transmitted on its own, 
\begin{eqnarray}\label{a6z}
T_{nn}=\int |T(p)|^2|A_n(p)|^2dp\equiv w_n.
\end{eqnarray}
\newline
We note that even initially uncorrelated particles may become correlated as a result of scattering.
This would happen, for example, if each transmitted one-particle state is significantly broadened in the coordinate space 
(narrowed in the momentum space), so that the integrals in Eq.(\ref{a6}) do not vanish, even if the integrals in Eq. (\ref{a4}) did.
\section {The generating function and full counting statistics}
For $N$ identical particles, there are $N+1$ outcomes, with $n=0,1,\ldots,N$ 
particles crossing in the barrier whose probabilities, $W(n,N)$, 
we will study next.
It is convenient to construct a generating function $G(\alpha)$,
\begin{eqnarray}\label{b1}
G^\pm(\A) =\lim_{t\to \infty}\la \Psi_{\text{out}}(t)|\Psi(t,\alpha)\ra
\ \ ,  
\end{eqnarray}
where
$|\Psi(t,\A)\ra \equiv K^{-1/2}\prod_{n=1}^N[\A t_n^+(t)+r_n^+(t)]|0\ra$,  
and $K$ is defined by Eq.(\ref{a8}).
By Wick's theorem, we have 
%
\begin{eqnarray}\label{b3}
G^\pm(\A) = S^\pm[\Delta^\pm_{mn}]/S^{\pm}[I_{mn}]\ \ ,  
\end{eqnarray}
where the matrix $\Delta$ is given by
\begin{eqnarray}\label{b4}
\Delta^\pm_{mn}
=I_{mn}+(\A-1)T_{mn},  \q n,m=1,2,...,N.
\end{eqnarray}
For the mean number of transmissions, $\overline{n}_T\equiv\sum_{n=0}^Nn W(n,N)$, we have
\begin{eqnarray}\label{b7}
\overline{n}_T(N,t_1,t_2,...,t_N)=\partial_\A G^{\pm}(\A)|_{\A=1} =\n
\sum_{j=1}^NS^\pm[I^{(j)}_{mn}]/S^\pm[I_{mn}]\q\q
\end{eqnarray}
where $I^{(j)}_{mn}$ is the matrix obtained from $I_{mn}$ by replacing the elements of the $j$-th row, 
$I_{j1},...,I_{jN}$ with $ T_{j1},...,T_{jN}$. The full counting statistics of an $N$-particle process are evaluated by noting that 
%
\begin{eqnarray}\label{c1}
W^{\pm}(n,N,t_1,t_2,...,t_N)  = \frac{1}{n!}\partial^n_\A G^{\pm}|_{\A=0}\q\q
\n
\ =\ \sum_{j_1<j_2<..<j_n}^NS^\pm[ I^{(j_1,j_2,...,j_n)}_{mn}]/S^\pm[I_{mn}]
\end{eqnarray}
%
where $ I^{(j_1,j_2,...,j_n)}_{mn}$ is the matrix obtained from $R_{mn}$ by replacing the elements of the rows
$j_1, j_2, ..., j_n$,  with the corresponding rows of the matrix $T_{mn}$.
Next we briefly discuss what would happen if the particles were not identical.
\section {The distinguishable particles limit}
The appearance of correlations between initially uncorrelated particles can be explained in the following way.
The particles are well separated initially, and if they leave the scatterer quickly enough, each scattering event occurs independently.
If, on the other hand, the scatterer detains each particle for a significant period of time, Bose or Fermi statistical effects become important while several (or all) particles are still inside. This, in turn, may alter measurable probabilities for various outcomes, which is the effect we seek to describe here.  
This is not possible if the particles are distinguishable, and do not interact with each other. Such particles cannot "meet" in the scatterer (or anywhere else), are always transmitted independently, and it does not matter if they arrive all at the same time, or their arrivals are separated by long time intervals.
\newline
Assume, for simplicity, that the individual tunnelling probabilities in Eq.(\ref{a6z}) are equal for all particles, $w_i=w_j\equiv w$. Then the probability for $n$ out of $N$ distinguishable particles (DP) to be transmitted is given by the binomial distribution, 
\begin{eqnarray}\label{dp1}
W^{DP}(n,N) =C^N_nw^n(1-w)^{N-n},   
\end{eqnarray}
where $C^N_n$ is the binomial coefficient. The mean number of transmissions, also independent of the choice of $t_1,t_2,...,t_N$, is given by 
\begin{eqnarray}\label{dp2}
\overline{n}_T^{DP}(n,N) =\sum_{i=1}^N w_i=wN.   
\end{eqnarray}
It is readily seen that the DP limit (\ref{dp1}) is reached if identical particles arrive at the scatterer after long intervals, 
$|t_i-t_j|\to \infty$. Indeed, in this limit all operators in Eqs.(\ref{a2}) and (\ref{a5}) commute, 
both $I_{mn}$ and $T_{mn} $ are diagonal, and for $w_i=w_j\equiv w$, Eqs. (\ref{c1}) yield
\begin{eqnarray}\label{dp1}
W^{\pm}(n,N,t_1,t_2,...,t_N)\to W^{DP}(n,N).   
\end{eqnarray}
We are, however, more interested in the case where quantum statistical effects do lead to measurable changes in in the channel probabilities $W^{\pm}(n,N,t_1,t_2,...,t_N)$, and will consider it next. 
\section {The two-particle case (N=2)}
In the simplest case of just two particles, $N=2$, 
Eqs.~(\ref{b7}) and (\ref{c1}) yield
%
\begin{align}
\label{d2}
\nonumber
W^\pm(2,2)\ = & \ \frac{w_1w_2\pm |T_{12}|^2}{1\pm|I_{12}|^2}\ \ ,
\\
W^\pm(1,2)\ = & \ \frac{w_1(1-w_2)+w_2(1-w_1)\pm 2\text{Re}(T_{12}R_{12}^*)}
                                                    {1\pm|I_{12}|^2}\ \ ,
\n
\overline{n}_T\ = & \ \frac{
        w_1+w_2\pm2 \text{Re}[T_{12}I^*_{12}]}{1\pm|I_{12}|^2}\ \ ,
\q\q \q\q\q\q\q
\end{align}
which coincides with the results of \cite{Baskin2014} if the particles have the same momentum distribution, $A_1(p)=A_2(p)$.
\newline
The last of the Eqs.(\ref{d2}) shows that if one sends to the scatterer initially correlated pairs of identical particles, ($I_{12}\ne 0$), the mean number of transmissions per pair may be different from the result obtained for two distinguishable particles in the same wave packet states.
\newline
If the pairs are not initially correlated, we always have $\overline{n}_T\ne \overline{n}^DP_T$, 
but the bosons (fermions) are more (less) likely to exit the scatterer on the same side.
This behaviour will be observed if the two particles, initially well separated from each other, 
meet in the scatterer, so that  $T_{12}\ne 0$.
\section {The $N$-particle case}
Equations (\ref{b7}) and (\ref{c1}) show that the results of the previous Section also hold for an arbitrary number of particles, $N>2$.
The mean number of transmissions 
may be affected by the symmetry of the initial state, $\overline{n}_T\ne\sum_{j=1}^Njw_j$, if, and only if the particles are correlated initially, $I_{mn}\ne \delta_{mn}$. 
\newline
For initially uncorrelated particles, $I_{mn}=\delta_{mn}$,  the symmetry changes 
 the probabilities $W(n,N)$, but not $\overline{n}$,  provided $T_{mn}\ne w_n\delta_{mn}$. In this case, ``bunching" and ``anti-bunching" types of behavior can be observed for bosons and fermions in the probability for all $N$ particles to be transmitted, $W(N,N)$. Since the matrix $T_{mn}$ is positive definite, the 
Hadamard inequality for determinants \cite{HornJohnson}, and its analog 
for permanents~\cite{Marcus1964} ensure that 
$W_{\pm}(n,N)^>_<\prod_{i=1}^Nw_i$ (cf.~also~Ref.~\cite{Pons2012}). 
Thus, $N$ bosons (fermions) are more (less) likely to be transmitted all together than DPs in the  same one-particle states.
Note that this argument cannot be extended to
 the probabilities $W(n<N,N)$, or for initially correlated initial states, $I_{mn}\ne \delta_{mn}$.
\newline
\section {Resonance transmission}
 A system likely to show these effects is a resonance barrier, where, due to the long delay in traversing it, even the particles well separated initially have a chance to ``pile up" in the scatterer. Transmission coefficient of such a barrier can be written as a sum of narrow Breit-Wigner peaks, 
\begin{eqnarray}\label{24}
|T(p)|^2\ =\ \sum_{l}
           \frac{\Gamma_l^2}{(p^2/2\mu-E^r_l)^2+\Gamma_l^2},
\end{eqnarray}
and even for initially uncorrelated particles, the shape of $|T(p)|^2|A(p)|^2$ may be narrow enough to ensure that 
$T_{mn}$ is not diagonal even if $I_{mn}=\delta_{mn}$.
\newline Figure 2 shows the mean number of transmissions for $N$ particles emitted after equal intervals, $|t_m-t_n|=T$, in identical Gaussian states of a coordinate width $\sigma$ and a mean momentum $p_0$, 
\begin{eqnarray}\label{24}
A(p)\equiv A_n(p)=(\sigma^2/2\pi)^{1/4}\exp[(p-p_0)^2\sigma^2/4]. 
\end{eqnarray}
The scatterer supports two resonant metastable states with the energies $E^r_{1,2}$ and widths $\Gamma_{1,2}$ of which one, or both can be accessed by the incident particle, as shown in the insets in Figs. 2 a and b.
With only one level involved, the mean number of transmissions $\overline{n}_T(T)$ for bosons raises to a maximum value for some correlated initial state (illustrated in Fig. 3 together with its fermionic counterpart), and 
\begin{figure}
	\centering
		\includegraphics[width=8cm,height=6cm]{{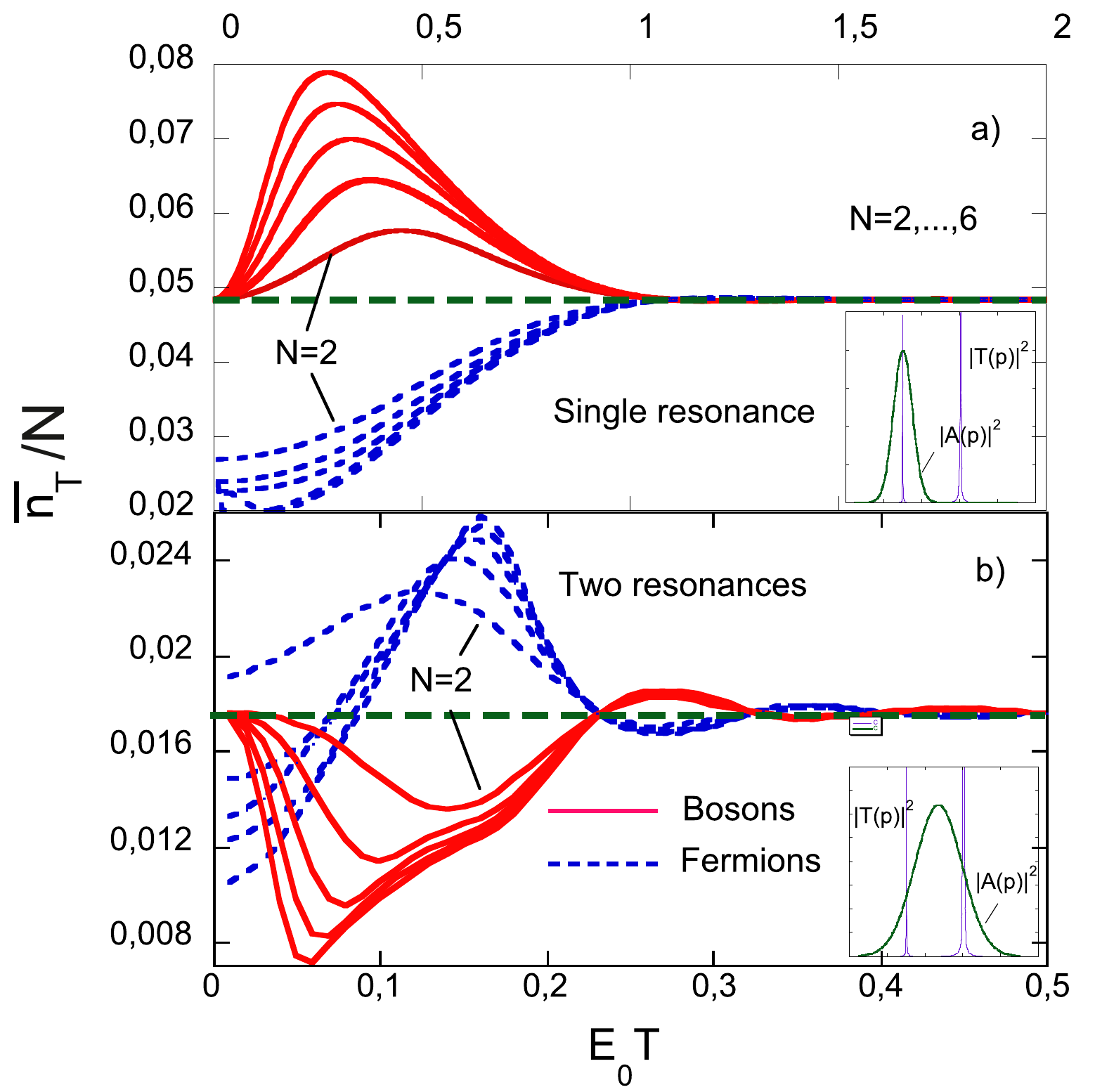}}
\caption{(Color online) Mean number of transmissions for $N$ particles in identical Gaussian wave packets
 vs. time $T$ between the emissions:
(a) through a single resonance level $E^r_1/E_0=0.41$, $\Gamma_1/E_0=0.0087$, $p_0\sigma=3.77$, and $E_0\equiv p_0^2/2\mu$ ; (b) through two resonance levels $E^r_1/E_0=0.95$, $\Gamma_1/E_0=0.038$, $E^r_2/E_0=3.82$, $\Gamma_2/E_0=0.28$ and $p_0\sigma=6.04$. Horizontal dashed lines correspond to the limit of distinguishable particles.
The momentum distribution $|A(p)|^2$ and the transmission coefficient $|T(p)|^2$ are shown in the insets.}

\label{fig:1}
\end{figure}
\begin{figure}
	\centering
		\includegraphics[width=8cm,height=3cm]{{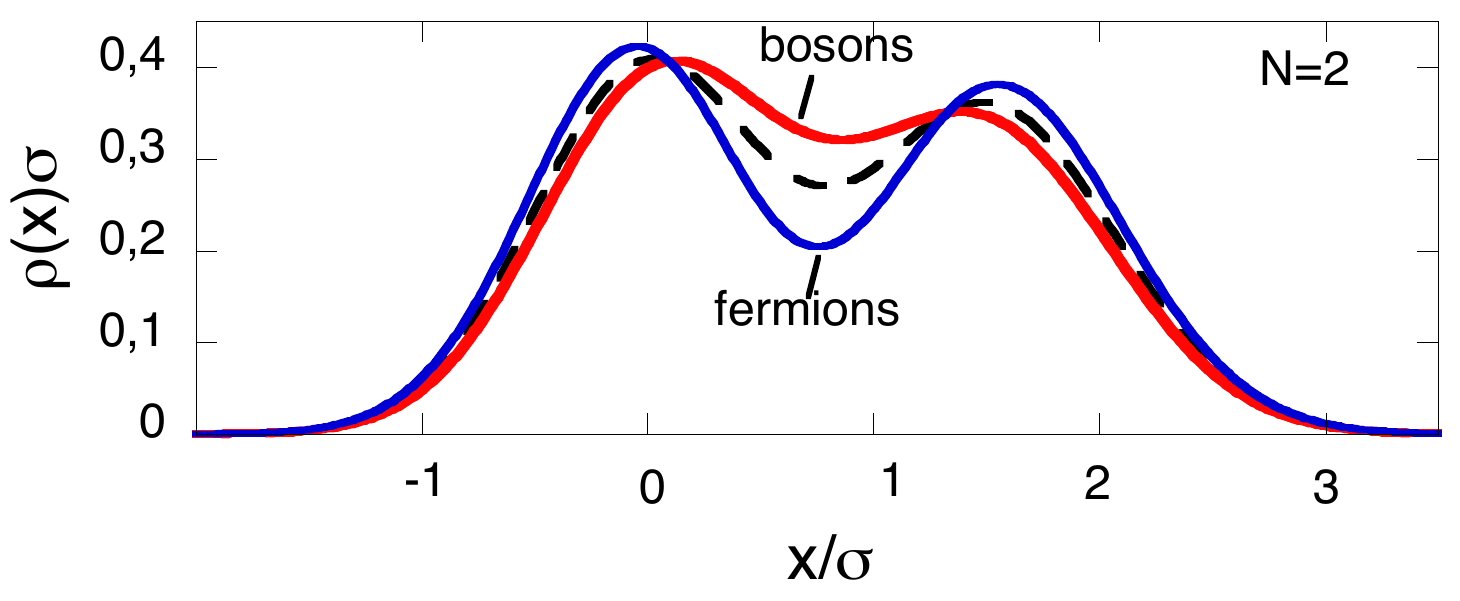}}
\caption{(Color online).One-particle density $\rho(x)$ (normalized to 2) for initially correlated 2-particle Gaussian state, $A_1(p)=A_2(p)\sim \exp[-(p-p_0)\sigma^2/2]$, $p_0\sigma=6$, $p_0T/\mu \sigma=1.5$, $p_0^2t/2\mu=4.5$, and $T\equiv t_2-t_1$.
 Also shown by the dashed line is $\rho(x)$ for two distinguishable particles in the same one-particle states. }
\end{figure}
then returns to the DP limit for initially uncorrelated particles (see Fig.2a). For fermions, the Pauli principle mostly reduces $\overline{n}_T$ to levels below the DP level, which for the maximally correlated states, obtained as $T\to 0$ \cite{FOOT}, is considerably reduced.
\newline
With two metastable states involved, interference between resonances  reverses the effect: for $0.1<p_0^2/2\mu T<0.25$, $\overline{n}_T$ is suppressed for bosons, and enhanced for fermions (see Fig.2b). 
\newline
The scattering probabilities $W(n,N,T)$ for the single resonance case, plotted in Fig.4 for $N=4$, show smooth deviations from the 
DP values in Eq.(\ref{dp1}) before reaching those values as the times between arrivals tends to infinity. We note that the probabilities 
$W(N,N)$ never fall below (exceed) their DP level for bosons (fermions), as discussed in the previous Section. We note also that
the increase or decrease in $\overline{n}_T$ results from a similar increase or decrease in the probability of the one-particle transmission channel, $W(1,N)$.
\begin{figure}
	\centering
		\includegraphics[width=8.2cm,height=5.8cm]{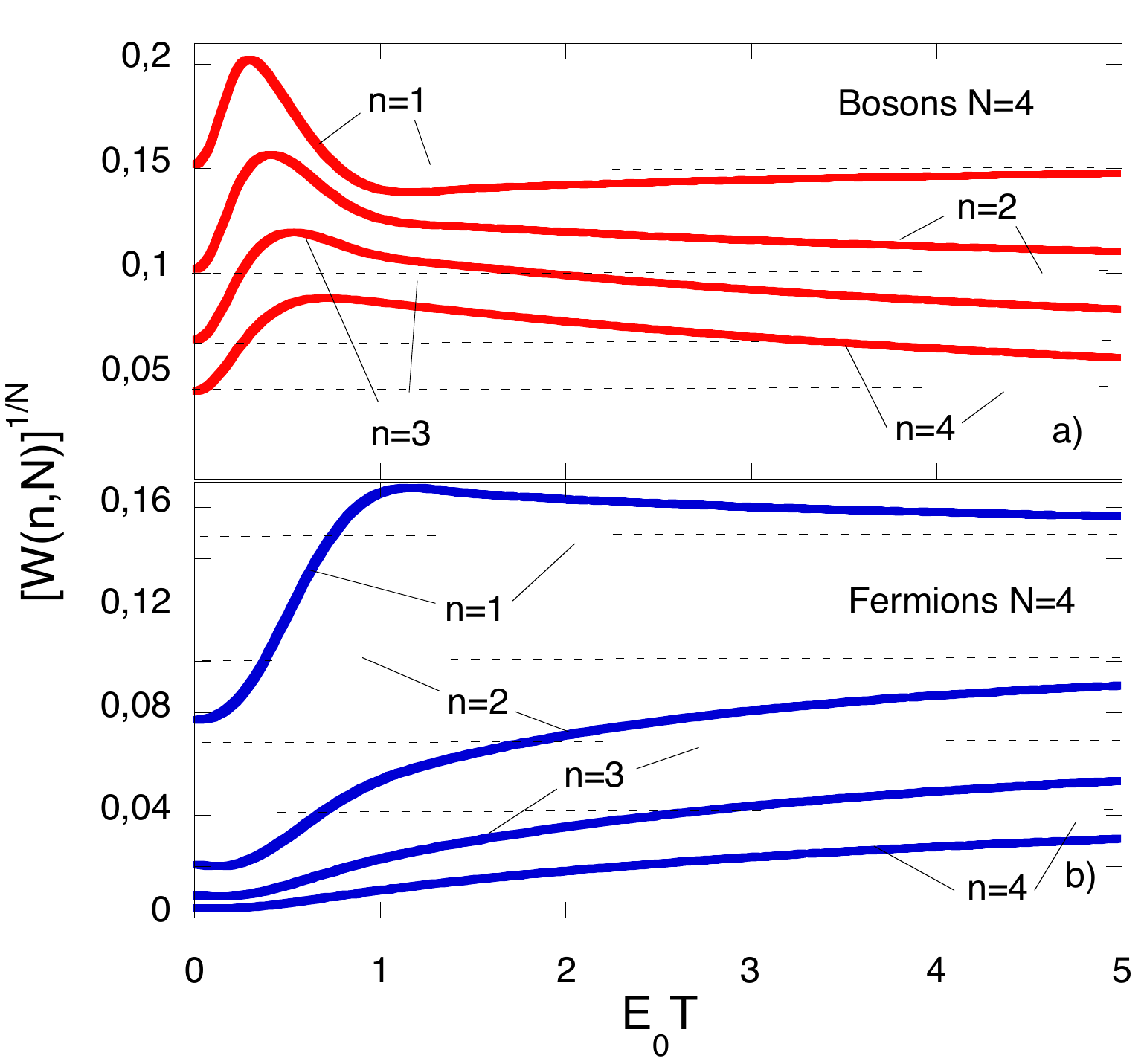}
\caption{(Color online) a)Probabilities for n=1,2,3,4 bosons to be transmitted for $N=4$ (single resonance); b) same as a), but for fermions. The parameters are as in Fig. 2a. Horizontal dashed lines indicate the the corresponding values for distinguishable particles
given by Eq.(\ref{dp1})
Incident particles may be considered uncorrelated for $E_0T\gtrsim 0.4$.}
\label{fig:1}
\end{figure}

With two resonances accessible to the particles,  the picture is more interesting. For bosons, $W(n,N,T)$ exhibit maxima, whenever the time between the emissions 
coincides with a multiple of the difference of resonant energies, 
$T\approx T_k= 2\pi k/(E^r_2-E^r_1)$, $k=1,2,\ldots$
The peaks are most pronounced for the (N,N) channel (cf. Fig. 5a) and, 
as shown in Fig.6a, become sharper as $N$ increases. This is another consequence of the symmetrization of the initial state which, with each particle distributed between the wave packets in Fig.1, appears to produce quasi-periodic excitation of the metastable two-level system supported by the barrier. With the number of particles increasing, the excitation looks more periodic, and the 
``resonance" condition $T\approx T_k$ needs to be satisfied with ever greater accuracy. Note that a similar (yet not identical) interference effects have been predicted for scattering trains of wave packet modes representing a single particle (for details see \cite{CAT}).
\begin{figure}
	\centering
		\includegraphics[width=8cm,height=6cm]{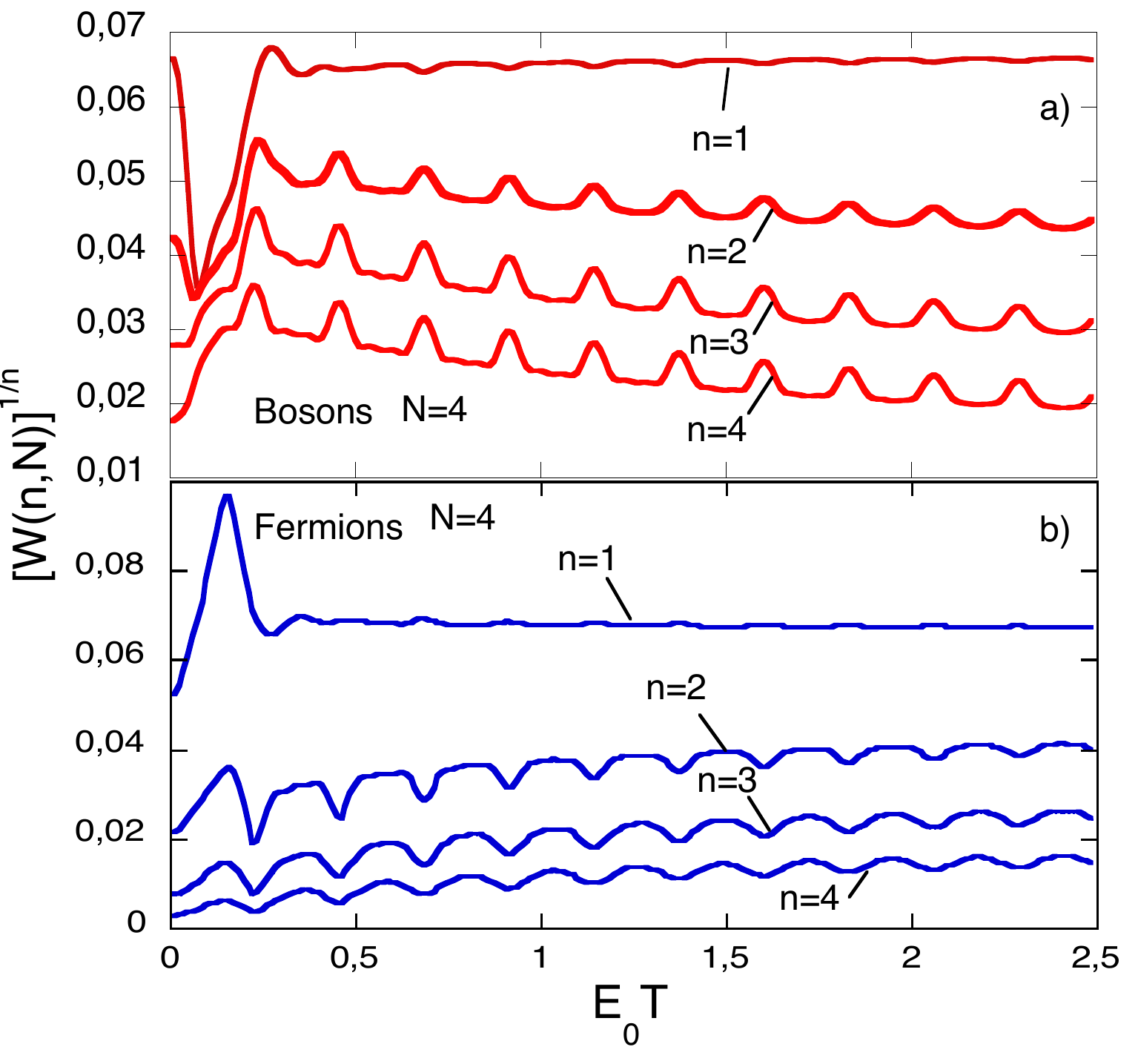}
\caption{(Color online) a)Probabilities for n=1,2,3,4 bosons to be transmitted for $N=4$ (two resonances); b) same as a), but for fermions. The parameters are as in Fig. 2b. Incident particles may be considered uncorrelated for $E_0T\gtrsim 0.4.$}
\label{fig:1}
\end{figure}
\newline
For fermions, probing two resonance states, the peaks at $T=T_k$ are replaced by dips, which appear, for example in the probability $W(2,4)$ shown in Fig. 5b.
 In contrast to the bosonic case, 
these dips are never seen in the $(N,N)$ channel, where $W(N,N,T)$ undergoes sinusoidal oscillations, no matter how large is the  numbers of particles $N$ (see Fig. 6b). 
\begin{figure}
\centering
		\includegraphics[width=8cm,height=6cm]{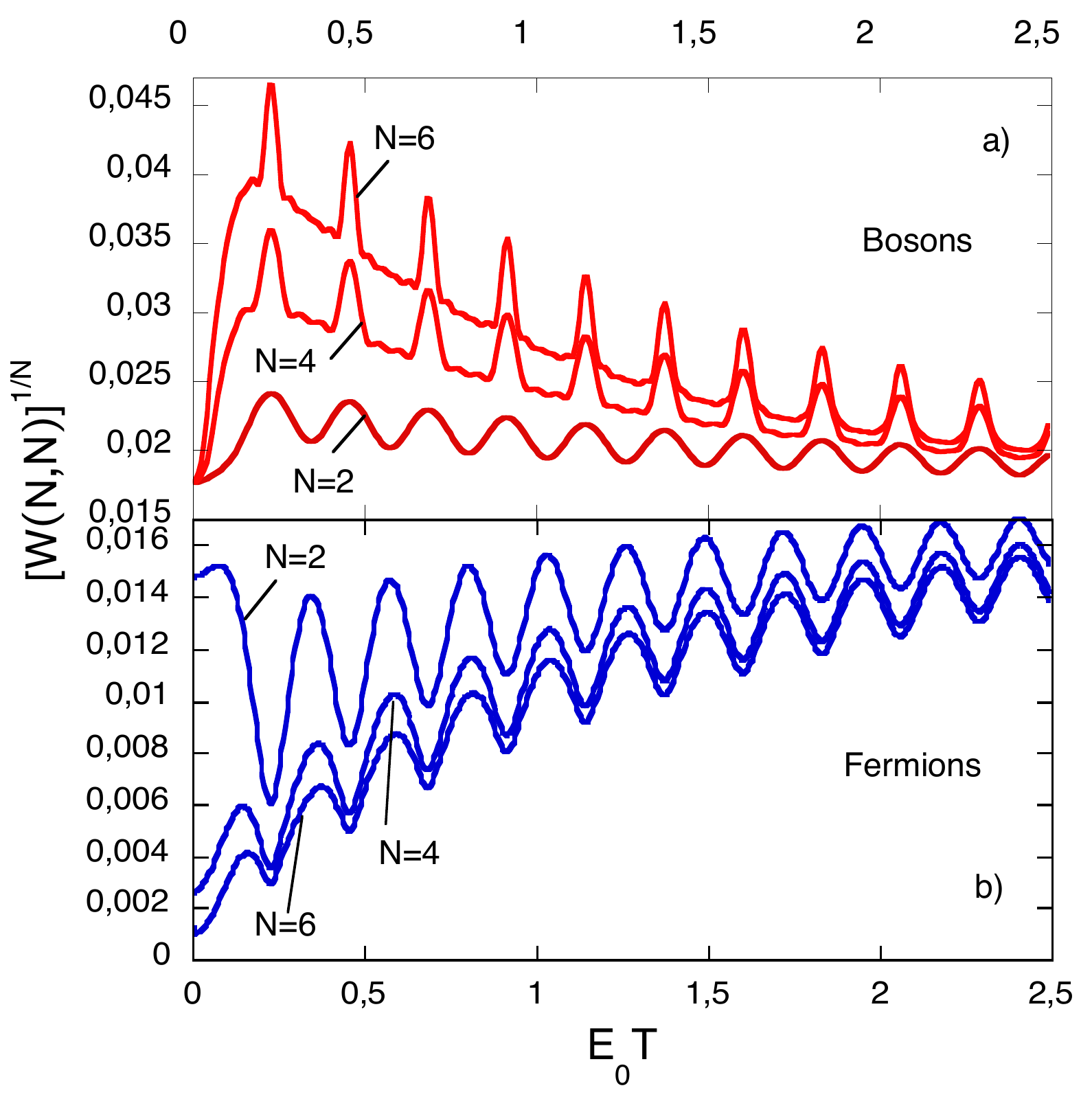}
\caption{(Color online) a)Probabilities for all bosons to be transmitted for $N=2,4,6$ (two resonances); b) same as a), but for fermions. The parameters are as in Fig. 2b.}
\label{fig:2}
\end{figure}
Experimental observation of effects of the Pauli principle on resonance tunneling would be possible for cold atoms in the Tonks-Girardeau regime, injected into quasi-one-dimensional trap with laser-induced barriers  \cite{RAIZ}.  
An optical realization of the bosonic experiment would consist in sending identically polarized photons toward a Fabry-Perot interferometer, or injecting them in a waveguide with narrowing, imitating a one-dimensional barrier. 
If required, a correlated initial state can be produced by scattering several uncorrelated particles off a long-lived resonance, and selecting the outcome in one of  the $n$-particle transmission channels. 
\newline 
\section{Conclusions and discussion}
In summary, Bose or Fermi statistics can significantly change the scattering outcomes 
for a train of non-interacting identical particles impinged on the same side of a scatterer. 
Physically, the effect requires simultaneous presence of several particles inside the scatterer. 
This  may occur if the initial state is already correlated, or if the particles, well separated initially, "pile up" inside, as a result of a scattering delay. In the latter case, the statistics of the process are affected in such a way that the mean number of transmissions per train, $\overline{n}_T$, remains unaffected. For a correlated initial state, $\overline{n}_T$ may be larger or smaller than that for the train composed of distinguishable non-interacting particles.
\newline
Mathematically, this is an interference phenomenon arising from the presence of additional terms in the (anti-) symmetrized wave function, which disappears if the particles can be distinguished.  Its analysis  is extremely simple, owing to the commutation of the evolution and symmetrisation operators: it is sufficient to first solve the corresponding one-particle problems, and then evaluate the asymptotic exchange integrals as $t\to \infty$.  Interference plays the most fundamental role in quantum mechanics, and we think it unlikely that a more detailed or more "physical" explanation of the effect can be provided.
\newline
To conclude, scattering of trains of identical particles
offers a variety of interference effects, very different from those observed in the HOM experiments, some of which were discussed in detail in Sect. VIII.  Observation of such effects is within the capability of modern experimental techniques.
\section{Acknowledgements:} 
We acknowledge support of the Basque Government (Grant No. IT-472-10), and the Ministry of Science and Innovation of Spain (Grant No. FIS2012-36673-C03-01). LB acknowledges the Russian Fund of Fundamental Investigations (Grant 12-01-00247) and the Saint Petersburg State University (Grant 11.38.666.2013) 


\newpage
 \begin{thebibliography}{999}
%
\bibitem{Aaronson2013}
  S.\ Aaronson and A.\ Arkhipov, Theory of Computing {\bf 9} , 143 (2013). 
%
\bibitem{HOM1987}
  C.K.\ Hong, Z.Y.\ Ou, and L.\ Mandel, Phys.\ Rev.\ Lett.\
{\bf 59}, 2044 (1987).
%
 \bibitem{DS} D. Sokolovski, Phys. Rev. Lett, {\bf 110}, 115302 (2013).
%
\bibitem{Sun2009}
 F.W.\ Sun and C.W.\ Wong, Phys.\ Rev.\ A {\bf 79}, 013824 (2009).
%
\bibitem{Shih1988}
Y.H.\ Shih and C.O.\ Alley,
Phys.\ Rev.\ Lett.\ {\bf 61}, 2921 (1988).
%
\bibitem{Gisin2007}
M.\ Halder, A.\ Beveratos, N.\ Gisin, V.\ Scarani, C.\ Simon, and H.\ Zbinden, 
Nat.\ Phys.\ {\bf 3}, 692 (2007).
%
\bibitem{Zeilinger2004}
P.\ Walther, J.-W.\ Pan, M.\ Aspelmeyer, R.\ Ursin, S.\ Gasparoni, and 
A.\ Zeilinger, Nature {\bf 429}, 6988 (2004).
%
\bibitem{mfb1}
J.G.\ Rarity and P.R. Tapster, J. Opt. Soc. Am. B. {\bf 6}, 1221 (1989).
%
\bibitem{mfb2}
Z. Y. Ou, J.-K. Rhee, and L. J. Wang,
Phys. Rev. Lett. {\bf 83}, 959 (1999).
%
%
\bibitem{mfb3}
X.-L. Niu, Y.-X. Gong, B. Liu, Y.-F. Huang, G.-C. Guo, and Z. Y. Ou, 
Opt. Lett, {\bf 34}, 1297 (2009).
%
\bibitem{TichyPRL2010}
M.C.\ Tichy, M.\ Tiersch, F.\ de Melo, F.\ Mintert, and
A.\ Buchleitner, Phys.\ Rev.\ Lett.\ {\bf 104}, 220405 (2010).
%
\bibitem{Mayer2011}
K.\ Mayer, M.C.\ Tichy, F.\ Mintert, T.\ Konrad, and
A.\ Buchleitner, Phys.\ Rev.\ A {\bf 83}, 062307 (2011).
%
\bibitem{SpagnoloPRL2013}
N.\ Spagnolo, C.\ Vitelli, L.\ Sansoni, E.\ Maiorino, P.\ Mataloni
{\em et al.}, Phys.\ Rev.\ Lett.\ {\bf 111}, 130503 (2013).
%
\bibitem{TG} M.\  Girardeau, J.\ Math.\ Phys. {\bf 1}, 516 (1960).
%
\bibitem{AspectPRA2013}
M.\ Bonneau, J.\ Ruaudel, R.\ Lopes, J.-C.\ Jaskula, A.\ Aspect,
D.\ Boiron, and C.I.\ Westbrook, Phys.\ Rev.\ A {\bf 87},
061603(R) (2013).
%
\bibitem{Aspect2015}
R.\ Lopes, A.\ Imanaliev, A.\ Aspect, M.\ Cheneau, D.\ Boiron, 
and C.I.\ Westbrook, e-print arXiv:1501.03065 (2015).
%
\bibitem{Leuenberger2014}
M.A.\ Khan and M.N.\ Leuenberger, 
Phys.\ Rev.\ B {\bf 90}, 075439 (2014).
%
\bibitem{Oriols2014}
D.\ Marian, E.\ Colom{\'e}s, and X.\ Oriols,
e-print arXiv:1408.1990 (2014). 
%
\bibitem{Urbina2014}
J.-D.\ Urbina, J.\ Kuipers, Q.\ Hummel, and K.\ Richter,
e-print arXiv:1409.1558 (2014).
%
\bibitem{Zurn2012}
G.\ Z{\"u}rn, F.\ Serwane, T.\ Lompe, A.N.\ Wenz, M.G.\ Ries, J.E.\
Bohn, and S. Jochim, Phys.\ Rev.\ Lett.\ {\bf 108}, 075303 (2012).
%
\bibitem{Rontani2012}
M.\ Rontani, Phys.\ Rev.\ Lett.\ {\bf 108}, 115302 (2012).
%
\bibitem{HunnPRA2013}
S.\ Hunn, K.\ Zimmermann, M.\ Hiller, and A.\ Buchleitner, 
Phys.\ Rev.\ A {\bf 87}, 043626 (2013).
%
\bibitem{Hassler2008}
F.\ Hassler, M.V.\ Suslov, G.M.\ Graf, M.V.\ Lebedev, G.B.\ Lesovik,
and G.\ Blatter, Phys.\ Rev.\ B {\bf 78}, 165330 (2008).
%
\bibitem{Baskin2014} 
D.\ Sokolovski and L.M.\ Baskin,  
Phys.\ Rev.\  A, {\bf  90}, 024101 (2014).
%
\bibitem{CAT} 
D.\ Sokolovski,  
Phys.\ Rev.\  A, {\bf  91}, 052104 (2015). 
%
\bibitem{FOOTm} For a photon in a waveguide $\mu$ is the effective mass acquired as a result 
of transversal confinement.
%
\bibitem{Loudon1998}
R.\ Loudon, Phys.\ Rev.\ A {\bf 58}, 4904 (1998).
%
\bibitem{FOOT1}Thus, $S^+[I_{mn}]$ is the permanent  ($\text{per}$) of the matrix $I_{mn}$, and $S^-[I_{mn}]$ is its determinant ($\det$).
%
\bibitem{FOOT2} One must also check that $T_{mn}\ne \text{const.}\times I_{mn}$.
%
\bibitem{FOOT} 
Note that for fermions, 
     $|\Psi_{in}\ra$ in Eq.(\ref{a71}), always normalized to unity 
     tends to a finite limit as $t_m-t_n\to 0$.  For example, for 
     two atoms prepared in Gaussian states of a width $\sigma$, 
     the one-particle density $\rho(x)$ tends to 
     $\exp(-2x^2/\sigma^2)[1+4x^2/\sigma2]$ as $t_2\to t_1$.
We study the scattering of this maximally correlated fermionic state, $I_{mn}=1$, whenever the limit $t_m-t_n\to 0$ is taken
in the rest of the paper.
%
\bibitem{HornJohnson}
R.A.\ Horn and C.R.\ Johnson, Matrix Analysis (Cambridge University
Press, New York, 1986).
%
\bibitem{Marcus1964}
M.\ Marcus, Proc.\ Am.\ Math.\ Soc.\ {\bf 15}, 967 (1964).
%
\bibitem{Pons2012}
M.\ Pons,\ D.\ Sokolovski, and A.\ del Campo,
Phys.\ Rev.\ A {\bf 85}, 022107 (2012).
%
\bibitem{RAIZ} T. P. Mayrath, F. Schreck, J. L. Hanssen, C. S. Chuu, and M. G. Raizen, 
Phys. Rev. A \textbf{71}, 041604 (2005).
\end{thebibliography}
\end{document}